# Near-Equilibrium Measurement of Quantum Size Effects Using Kelvin Probe Force Microscopy

Thomas Späth,[†] Matthias Popp, Carmen Pérez León*, Michael Marz, and Regina Hoffmann-Vogel,[‡]

Physikalisches Institut, Karlsruhe Institute of Technology (KIT),
Wolfgang-Gaede-Str. 1, 76128 Karlsruhe, Germany

**ABSTRACT:** In nanostructures such as thin films electron confinement results in the quantization of energy levels in the direction perpendicular to the film. The discretization of the energy levels leads to the oscillatory dependence of many properties on the film thickness due to quantum size effects. Pb on Si(111) is a specially interesting system because a particular relationship between the Pb atomic layer thickness and its Fermi wavelength leads to a periodicity of the oscillation of two atomic layers. Here, we demonstrate how the combination of scanning force microscopy (SFM) and Kelvin probe force microscopy (KPFM) provides a reliable method to monitor the quantum oscillations in the work function of Pb ultra-thin film nanostructures on Si(111). Unlike other techniques, with SFM/KPFM we directly address single Pb islands, determine their height while suppressing the influence of electrostatic forces, and, in addition, simultaneously evaluate their local work function by measurements close to equilibrium, without current-dependent and non-equilibrium effects. Our results evidence even-odd oscillations in the work function as a function of the film thickness that decay linearly with the film thickness, proving that this method provides direct and precise information on the quantum states.

## 1 Introduction

Electrons in atomically flat metallic thin films located on surfaces with a band gap are confined in the potential well formed between the vacuum level and the substrate band gap. The confinement results in the quantization of the energy levels[1-3]. In thin film nanostructures, i.e. islands, the confinement of the electrons takes place in the z-direction perpendicular to the surface, leading to a pronounced effect of the quantization of the perpendicular wavevector component $k_z$, while $k_x$ and $k_y$ remain quasi-continuous. Consequently, the Fermi sphere of allowed states is reduced to a discrete number of Fermi discs with a constant $k_z$ value[1,4] (see Figure 1a). If one were able to increase the film thickness continuously, the number of Fermi discs would grow and would move closer together in the reciprocal space[3] (see Figure 1a). The periodic addition of Fermi discs results in oscillations of the Fermi energy ($E_F$) i.e. of the chemical potential, as a function of the film thickness[4]. As a consequence quantum size effects (QSE) lead to an oscillatory behavior of many physical and chemical properties as a function of the film thickness, such as the surface energy[4], electric resistivity[5], the superconducting critical temperature[6,7], the surface chemical reactivity[8,9], the work function and accordingly the chemical potential[8,10,11]. The work function is defined as $\Phi = E_{vac} - \mu$, $E_{vac}$ being the energy of the vacuum level, and µ the chemical potential, see Figure 1e. Typical systems where QSE have been observed are Ag/Au(111)[2], Ag/Fe(100)[3], Pb/Cu(111)[12], Pb/Ag(111)[13], and Pb/Si(111)[4,6-11,14-16]. Among these systems, Pb on Si(111) has been intensely investigated, since it exhibits a rich variety of pronounced QSE[17].

In Pb thin films, the Fermi energy remains pinned to one Fermi disc until the number of electrons exceeds the number of states available in that disc, then it jumps to the next one. This occurs with good approximation when the thickness of the film a reaches

$$a = n\,(\lambda_F/\,2), \qquad (1)$$

where $\lambda_F$ is the Fermi wavelength that varies with the film thickness and $n$ an integer number. The thickness of an atomic Pb layer, $d_0 = 0.285$ nm[18], satisfies approximately the relation

$$2d_0 \approx 3\,(\lambda_{F,\text{bulk}}/\,2), \qquad (2)$$

where $\lambda_{F,\text{bulk}} = 0.394$ nm is the bulk Fermi wavelength of Pb[18-20]. Thus, if the film thickness is increased by two atomic layers, three states are added and the $E_F$ remains nearly constant. If, however, the film is increased by just one atomic layer, only one state is added and there is a strong difference in $E_F$. This yields to oscillations of $E_F$ and the work function with the film thickness. The amplitude of these oscillations decays as more and more Fermi discs become populated, i.e. the nanostructure gradually approaches bulk properties[4]. This behavior is shown in Figure 1b, where the graph has been generated with a simple model described in the Supporting Information (SI, Section I). The oscillation with a wavelength of $\lambda_{F,bulk}/2$ is additionally modulated by a quantum beating pattern that causes a periodical reversal of maxima and minima[19]. This beating pattern results from the interference of the damped oscillations and the discrete nature of the atomic layer structure of the film[4,19]. As indicated in Equation 2, the relation is not exactly 3 but rather $2d_0 \approx 2.9 \cdot (\lambda_{F,bulk}/2)$. Since the value 2.9 differs by 10% from the next integer 3, every ten atomic layers a phase slip is expected. The ultra-thin thickness limit of the first beating merits particular attention because in this regime the amplitude of the oscillations is the largest[4,10,19,20].

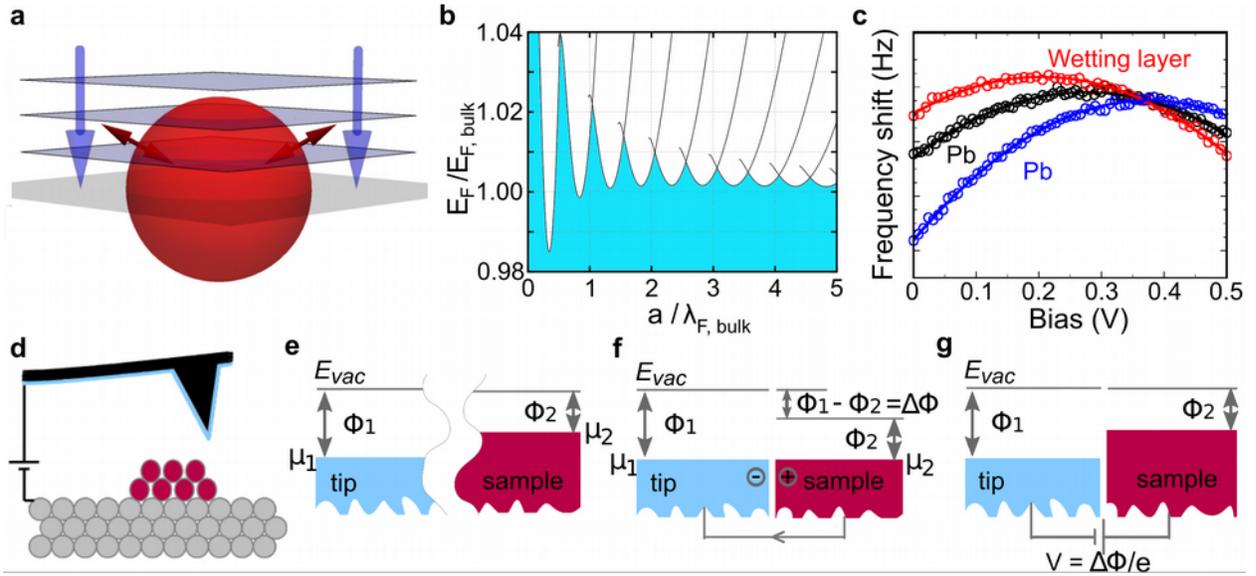

**Figure 1.** Quantum size effects on thin metal films and fundamentals of the Kelvin method. a) The Fermi sphere of allowed states is reduced to a discrete number of Fermi discs in $k_z$-direction ($n = 2$ in the sketch, see SI Section I). b) Oscillation of the Fermi energy of a Pb(111) film with the film thickness. The depth of the potential well is $1.05 \cdot E_{F,bulk}$, see SI Section I. c) Kelvin parabolas measured on different parts of the sample. d) Sketch of tip and sample configuration. e) Electronic energy levels of two materials at infinite distance from each other. f) Charge transfer when the two materials are brought together. g) Minimisation of electrostatic forces using the Kelvin method.

In order to detect the quantum oscillations in the work function two approaches were applied in the past. One approach was photoemission[2,17]. Since this method integrates over a large surface area, it integrates over all film thicknesses present within this area. The main difficulty is to produce a large fraction of thin films with a homogeneous thickness on the one hand, and to interpret the integrated data on the other hand. Another approach was to address single islands, using scanning tunneling microscopy (STM) with a lock-in technique. This was achieved by measuring the tunneling current as a function of tip-sample distance and bias voltage. Following, the results were fitted to models including the work function[8,10,11]. This method led to contradictory results, because the oscillations observed using STM varied in phase depending on the bias voltage applied, an effect attributed to the electronic state the tunneling electrons occupy in the Pb island[11].

Here, we show how simultaneous Kelvin probe force microscopy (KPFM, see Ref. 21) and scanning force microscopy (SFM) assist in explaining these results by directly measuring the influence of QSE on the local work function (LWF) in ultra-thin Pb islands on Si(111). The main advantage of this method vs. STM is that the measurement of the LWF is carried out very close to equilibrium, since KPFM determines the LWF without extracting electrons, i.e. essentially without any current. Thus, we suppress current-dependent and non-equilibrium effects.

With KPFM, the work function difference between tip and sample, $\Delta\Phi$, is directly determined during SFM measurements (Figure 1d). The microscope is operated in the dynamic frequency modulation mode, where the tip oscillates at its resonance frequency, and the frequency shift of this oscillation is a measure of the tip-sample interaction, i.e. of the force between them. Since the tip-sample system acts as a capacitor with a capacitance C, applying a bias voltage V leads to parabolic force F between tip and sample,

$$F = \tfrac{1}{2} (\partial C / \partial z) (V - \Delta\Phi/e)^2, \qquad (3)$$

where z is the tip-sample distance. The tip – Si or Pt/Ir-coated Si here – and the sample – Pb – have different WF (Figure 1e). When the tip is approached to the sample, charge transfer between tip and sample leads to a shift of the parabola by $V = \Delta\Phi/e$, and the chemical potentials align (Figure 1f). Figure 1c shows the parabolic depen-dence of the frequency shift on the bias voltage measured on different parts of the sample (see Equation 3). Applying a bias voltage between tip and sample that equals their WF difference, the Kelvin voltage ($V_{Kelvin} = \Delta\Phi/e$), electrostatic forces between them are minimized (Figure 1g). Assumptions about the shape of the potential well between tip and sample are not needed if the distance between tip and sample is large enough to avoid chemical bonding. At this distance range, the measured LWF does not depend on the tip-sample distance[22]. The KPFM measurements were performed in the range where the current flow was below the detection limit of our experimental setup.

## 2 Experimental Section

2.1 Sample preparation:

The n-type Si(111) crystals (phosphorous-doped, $\rho = 7.5$ Ωcm) were first shortly heated to 1200° C in a commercial ultra-high vacuum chamber (Omicron NanoTechnology GmbH, Taunusstein, Germany) with a base pressure below $3 \cdot 10^{-8}$ Pa. Prior to Pb evaporation, for some samples the Si(111) substrate was cooled down to room temperature, whereas for other samples the Si substrate was first cooled to liquid nitrogen temperatures and then annealed at room temperature for times between 20 and 40 min. After Pb deposition, the samples were transferred immediately to the liquid nitrogen pre-cooled scanning force microscope (Omicron VT-AFM) attached to the same vacuum vessel for the measurements.

2.2 Scanning Force Microscopy Methods:

For low temperature scanning force microscopy (SFM) imaging, we used commercial Si cantilevers (Nanosensors, Neuchatel, Switzerland) with a spring constant of 30−60 N/m and resonance frequencies of approximately 300 kHz. The majority of the tips were coated with a 25 nm thick layer of Pt/Ir. The tips were first degassed in vacuum at 150° C for several hours. Subsequently, they were cleaned by Ar sputtering. The tips were transferred to the scanning force microscope located in the same vacuum vessel and tested on clean Si(111). The cantilever was oscillated at resonance at a constant amplitude of $5 − 10$ nm. We used a Nanonis phase-locked loop electronics (SPECS, Zurich, Switzerland) to detect the frequency shift induced by the tip-sample interaction. We operated in the dynamic mode with frequency modulation technique (FM-SFM); this topographic detection mode is also called the non-contact mode. We performed Kelvin probe measurements in parallel to the topography measurements by applying an oscillating voltage to the tip with a typical frequency of $f_{Kelvin} = 266$ Hz and an amplitude of 0.5 V. The sign of the measurement has been calibrated using a KBr(001) surface[23]. Negative charge accumulates on alkali halide crystals. We represent this negative charge by larger work function differences (bright contrast), i.e. higher Kelvin voltage. This convention was also double-checked on the Si(111)-(7 × 7) reconstructed part of a stepped Si (7 7 10) surface, where the negative charge expected on the restatoms was also correctly reproduced[24].

## 3 Results and Discussion

When a clean Si(111) surface is covered with small amounts of Pb, (111)-oriented islands grow following a Stranski-Krastanov growth mode surrounded by a disordered wetting layer (WL)[14,16,25]. The islands with a flat-top mesa shape exhibit preferred heights influenced by the QSE[14,16,25]. At room temperature, thermally activated diffusion is strong and makes the islands grow with time. The islands' size can be controlled by varying the deposition parameters and annealing time. In previous STM studies, large Pb islands extending over many terraces were grown over stepped regions[7-11]. These flat-top mesas contained regions of different heights in one island. Our focus here is to concentrate on small islands covering only a few terraces, ideally only one. Thus, we avoid the influence of epitaxial strain of islands overgrowing many terraces due to the misfit of Pb(111) and Si(111) in the vertical direction[8]. We also aim at the limit of ultra-thin islands of only a few atomic layers in height, particularly below 10 atomic layers. In this height range, the amplitude of the quantum oscillations is the largest[4,10,19,20], as we mentioned above, and the island height measurement particularly precise. Hence, after Pb deposition, we cooled down the sample to ∼110 K by inserting it into the pre-cooled SFM to stop the diffusion process.

Figure 2 shows SFM topographic and simultaneously obtained Kelvin images of ultra-thin Pb islands on Si(111). For an atomically resolved image of a Pb island, see Figure S1 of the SI. The islands have a small lateral extension and heights in the desired range. The Kelvin images, Figure 2b,d, confirm that the LWF differs for islands of different heights. As already noticed in Figure 1c, also the LWF of the islands differs from the one of the WL. The lower Kelvin voltage values on the islands indicate electron depletion with respect to the WL, following the description that higher Kelvin voltage implies more negative charge[23,24,26]. These results are in contrast to a previous SFM study that found no Kelvin voltage difference within a large Pb island with different heights[27]. That study used a tuning fork sensor with a higher stiffness and less sensitivity to the electrostatic force than the cantilevers used here.

In Figure 2a, some islands cover more than one Si(111) terrace. Also on such islands regions with different heights show different work functions, as seen in Figure 2b. In Figure 2c, a few islands have a ring-like shape. In two of these islands in the middle of the image, the local island height measured from the WL is given in Pb atomic layers.

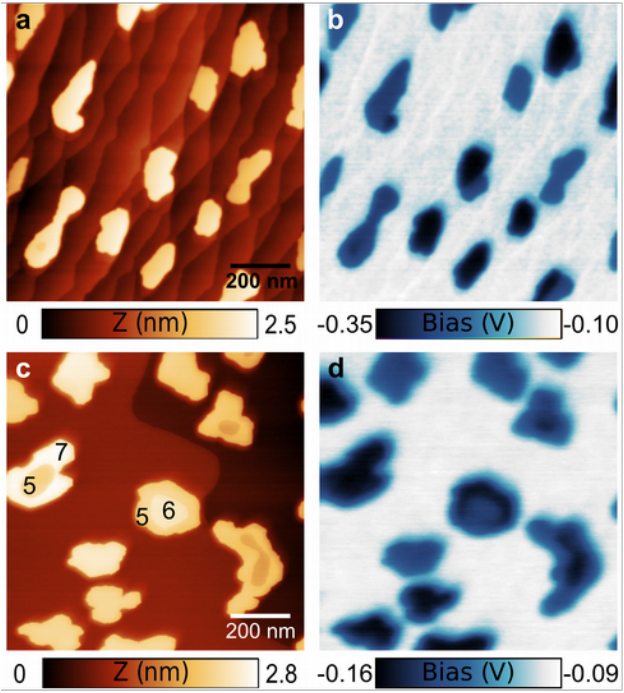

**Figure 2.** SFM topographic (a, c) and simultaneously obtained Kelvin images (b, d) of ultra-thin Pb islands on Si(111). The numbers indicate the local height in atomic layers measured from the WL. Islands having internal height differences of one atomic layer show differing LWF within the island, while for internal height differences of two atomic layers the LWF is nearly equal over the whole island. This underlines the bilayer period of the quantum oscillation. Images size (1 × 1) $\mu m^2$; frequency shift (a, b) $\Delta f = -30$ Hz, (c, d) $\Delta f = -20$ Hz; oscillation amplitude (a, b) $A = 5.3$ nm, (c, d) $A = 10$ nm; normalized frequency (a, b) $\gamma = -2.1$ fN$\sqrt{m}$, (c, d) $\gamma = 3.6$ fN$\sqrt{m}$. Pt/Ir coated tips.

These islands are particularly interesting, because they directly show the bilayer oscillation of the LWF on Pb islands[15]. For the 7 and 5 atomic layers high island, the whole island has a similar LWF, while for the 5 and 6 atomic layers high island, both parts display a strongly different LWF (see Figure 2d).

Another example of a Pb island covering several terraces is presented in Figure 3. Unlike many single crystalline islands that have a mesa shape with an uniform height at the top surface, this particular island is composed of three coalescent flat-top crystallites (labeled as *a*, *b* and *c* in Figure 3a). The height in atomic layers (measured from the WL) of the distinct individual parts of the island is indicated in Figure 3b,f (5 − 8 atomic layers). Due to the misfit of Pb(111) and Si(111) in the vertical direction[8], the *a* crystallite slightly overtops the *b* and *c* crystals, as observed in Figure 3a. In addition, this height difference is noticeable as a step of subatomic height (∼ 40 pm) at the top of the island in the schematic cross section presented in Figure 3f. Owing to the rich morphology of this island, we have used it for investigating the influence of the QSE on SFM measurements.

SFM topographic imaging of the surface has been performed with the Kelvin controller switched on (Figure 3c,e), and with the Kelvin controller switched off (Figure 3d). The contrast is adjusted in the images to enhance the height differences. The topographic image in Figure 3d obtained with the Kelvin controller switched off shows an island composed of many small crystals, these corresponding to the parts of the island indicated in Figure 3b that cover different terraces. However in Figure 3c imaged while minimizing electrostatic forces, the three crystallites that compose the island are well resolved. This demonstrates that the image in Figure 3d was affected by electrostatic forces due to the strong variations of the LWF in the distinct parts of the islands, which are notably prominent in Figure 3e.

It has been shown that to obtain reliable height measurements in samples with different materials or regions with differing electrostatic interactions the use of KPFM is mandatory[28]. Otherwise, the apparent height is strongly influenced by the tip-sample work function difference. Recent calculations have evidenced that owing to the QSE, electrostatic forces above islands are significant[9]. The quantization of the electronic states even influences the topographic STM measurement causing the imaging of buried interfacial structures, that may mislead the interpretation of the topographic images of the islands[10,29].

In Figure 3c,e is also noticeable that for parts of the islands having an odd number of Pb layers from the WL, the local Kelvin voltage is significantly lower than in parts of the islands having an even number of atomic layers. This again evidences the bilayer oscillation of the LWF as a function of island height.

In order to address the quantum oscillations in the LWF of ultrathin Pb islands on Si(111), we grew a large amount of small islands and investigated them with simultaneous SFM/KPFM, as in Figures 2,3. The obtained data were analyzed by applying statistical methods that we detail in the SI Section III. The particularity of our approach is that we use the WL as reference, instead of using the tip, the standard reference also in STM studies[11]. Consequently, in the analysis the height of the islands was measured from the WL. Likewise, we evaluated the LWF of the island in terms of the LWF of the WL. As we mentioned above, KPFM directly provides the work function difference between sample and tip:

$$e \cdot V_{Kelvin} = \Delta \Phi = \Phi_{sample} - \Phi_{tip} . \quad (4)$$

Using the work function of the WL as reference:

$$\begin{aligned}\Delta \Phi^* &= e \cdot V_{Kelvin, island} - e \cdot V_{Kelvin, WL} \\ &= \Phi_{island} - \Phi_{tip} - (\Phi_{WL} - \Phi_{tip}) \quad (5) \\ &= \Phi_{island} - \Phi_{WL} ,\end{aligned}$$

we obtain tip-independent results. Since the LWF of the WL is homogeneous within the experiments shown here (see SI Section IV), it only contributes

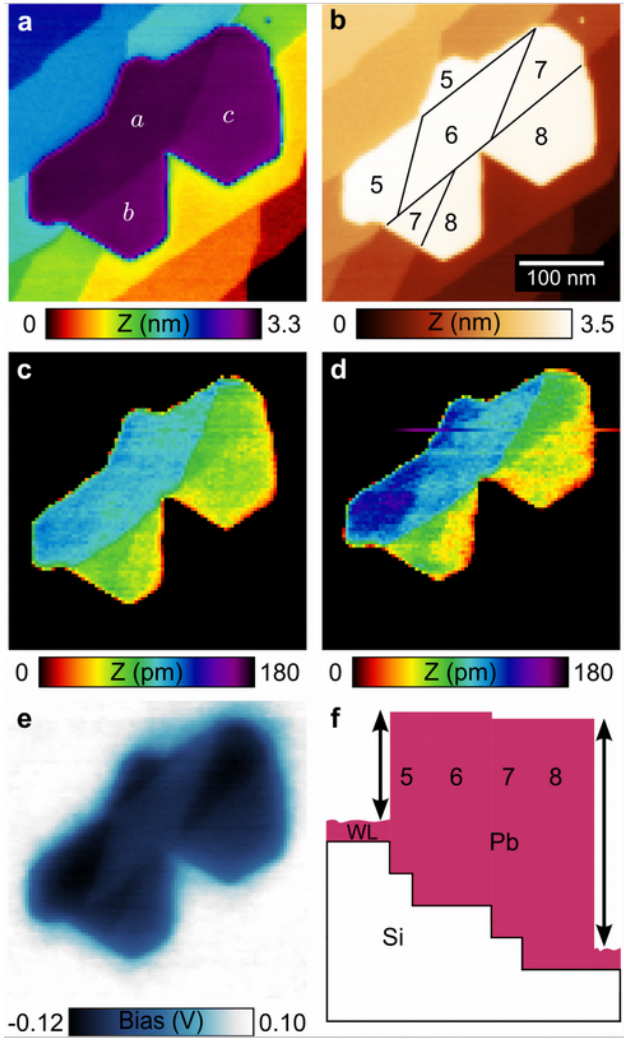

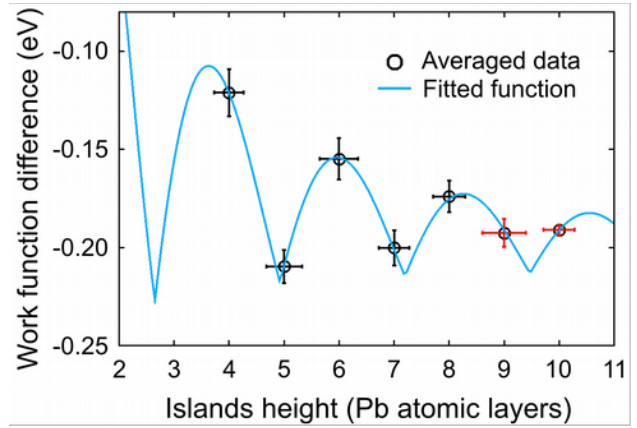

**Figure 4.** Quantum oscillations in the local work function of ultra-thin Pb islands on Si(111) as a function of the island height measured from the WL obtained with SFM/KPFM. The LWF exhibits a damped even-odd oscillation with increasing island height. The two points with red error bars result from a small number of data. Data fitted with Equation 6.

**Figure 3.** SFM (a–d) and Kelvin (e) images of a Pb island composed of three flat-top crystallites extending over several Si(111) terraces. The height of the island varies locally and consequently also the LWF. a) Crystallites labeled a, b and c. b) Local height of the island given in atomic layers measured from the WL. c) Topography obtained with the Kelvin controller switched on. d) Topography obtained with the Kelvin controller switched off. The image is influenced by electrostatic forces. e) Corresponding Kelvin voltage of image c. f) Schematic cross section of the island. Image size (350 × 350) nm$^2$; frequency shift $\Delta f = -17$ Hz; oscillation amplitude $A = 8$ nm; normalized frequency $\gamma = 2.1$ fN√m. Pt/Ir coated tip.

as a constant offset.

Figure 4 summarizes the evaluated data obtained with one single tip: the quantum oscillation of the LWF of Pb islands on Si(111) as a function of island height is directly obtained with SFM/KPFM. The amplitude of the oscillation decays for increasing island height, as expected. The advantage of using the WL as reference is that we can compare the results of different experiments directly. In Figure S4 of the SI results obtained with diverse tips align well with each other, confirming that we indeed get tip-indepen dent results. The relative change of the LWF from layer to layer, the phase of the beating pattern, and the decay of the oscillation are the most relevant quantities in this study, and these do not depend on the tip.

In the literature[4,15,17,20], the damped sinusoidal function,

$$f(N) = ((A|\sin Nk_F d_0 + \varphi | + B) / N^\alpha) + C, \quad (6)$$

has been used to describe the periodicity and decay of the oscillation of the surface energy and work function as a function of the number of atomic layers $N$. In Equation 6, $A$, $B$, $C$, $\varphi$, and $\alpha$ are $N$-independent constants, and $k_F$ is the Fermi wave vector measured from the zone boundary[20] (for details see SI Section V). To evaluate the decay exponent of the oscillations, we have fitted this equation to our data. The values obtained for the free parameters show that this description is suitable, see Table S1 of the SI. We find for the decay exponent $\alpha$ a value of 1.26. This value differs from the conventional $\alpha = 2$ of the Friedel oscillations in electron density at other metal surfaces[30], we obtain also this value in the simple model described in the SI Section I. However, as stated in Ref. 30, Fermi surface nesting leads to one-dimensional properties of the electrons in Pb(111) causing a deviation from simple physics that leads to a decay exponent $\alpha = 1$ for the work function.

In the introduction we announced the presence of a beating pattern. To study the behavior of the amplitude of the oscillations above 10 atomic layers, we grew larger islands (see details in SI Section IV). Our results presented in Figure S5 confirm a phase slip on the oscillations and show a knot around 9 − 10 atomic layers. Using these additional data, we fitted again $\alpha$ and obtained a value of 1.02 (see SI Section V).

So far an analysis of the work function oscillation as a function of island height has been mainly performed using STM with a lock-in technique[8,10,11]. The oscillation has been observed and predicted up to an island height of 32 atomic layers[10,18]. It is difficult to compare our data with previous STM works, since the phase of the oscillation varies from publication to publication reversing minimum and maximum. This is partly due to different ways to count the number of atomic layers, e.g., including or excluding the WL which is generally agreed to be around 1 atomic layer thick[16,31]. Another reason is the dependence of the STM-obtained LWF on the tunneling voltages used, as we mentioned above[11,13]. The maximum relative change from layer to layer in our SFM/KPFM studies is of $0.08 - 0.05$ eV, whereas in the literature for islands above 10 atomic layers, this change is of $0.2 - 0.15$ eV[8,10], although a smaller value is expected due to the decay of the oscillation. An additional point to compare with is the thickness at which the knot of the beating pattern occurs, i.e. the phase of the beating. Some authors observe the first knot around $9 - 10$ atomic layers[8,10], like we do here, while others observe it at around $12 - 13$ atomic layers[11]. This difference is hardly expected to be caused by different ways of counting the atomic layers or to imprecision by measuring the height. A possibility that may explain such difference could be differences in the metal-semiconductor interface that modify the position of the knots[31-34].

## 4 Conclusions

Summarising, we have demonstrated that the SFM/KPFM combination provides a reliable method for measuring the influence of QSE on the work function of metallic ultra-thin films. The advantages of this technique vs. previous approaches is that we address single films and unambiguously determine their local thickness by suppressing the influence of the QSE on the topographic measurements, while simultaneously evaluating their LWF by measurements close to equilibrium, i.e. essentially without any current. In addition, in our study we have used the strategy, on the one hand, of concentrating on small ultra-thin islands where the amplitude of the quantum function oscillation is the largest and the influence of strain low. And, on the other hand, of using the WL as reference, thus getting tip-independent results. Our results show a qualitative and quantitative agreement with previous results in the literature (oscillating behavior, decay constant, phase of the beating pattern). In conclusion, the SFM/KPFM method is well-suited to determine the quantum oscillation in the work function with utmost precision.

## 5 Acknowledgements

We thank H. v. Löhneysen and M. Weides for enlightening discussions. Funding: Financial support from the European Research Council through the Starting Grant NANOCONTACTS (No. ERC 2009-Stg 239838) and from the Ministry of Science, Research and Arts, Baden-Wuerttemberg, in the framework of its Brigitte-Schlieben-Lange program is gratefully acknowledged.

## ASSOCIATED CONTENT

Supplementary Information (SI) available: [Simple model for calculating the Fermi energy, atomically resolved SFM image, statistical averaging methods, extra data at large island heights, and fitting parameters. Figures S1–S5 and Table S1].

## AUTHOR INFORMATION

**Corresponding Author**

* E-mail: cperezleon.science@gmail.com

**Present Addresses**

† Present address: Department of Materials Science, Darmstadt University of Technology, Jovanka-Bontschits-Str. 2, 64287 Darmstadt, Germany
‡ Present address: Institut für Angewandte Physik, Karlsruhe Institute of Technology (KIT), Wolfgang-Gaede-Str. 1, 76128 Karlsruhe, Germany

## References

1 F. K. Schulte, *Surf. Sci.*, 1976, **55**, 427–444.
2 T. C. Chiang, *Surf. Sci. Rep.*, 2000, **39**, 181–235.
3 M. Milun, P. Pervan and D. P. Woodruff, *Rep. Prog. Phys.*, 2002, **65**, 99–141.
4 P. Czoschke, H. Hong, L. Basile and T.-C. Chiang, *Phys. Rev. B*, 2005, **72**, 075402.
5 M. Jałochowski and E. Bauer, *Phys. Rev. B*, 1988, **38**, 5272–5280.
6 Y. Guo, Y.-F. Zhang, X.-Y. Bao, T.-Z. Han, Z. Tang, L.-X. Zhang, W.-G. Zhu, E. G. Wang, Q. Niu, Z. Q. Qiu, J.-F. Jia, Z.-X. Zhao and Q.-K. Xue, *Science*, 2004, **306**, 1915–1917.
7 D. Eom, S. Qin, M. Y. Chou and C. K. Shih, P*hys. Rev. Lett.*, 2006, **96**, 027005.
8 X. Ma, P. Jiang, Y. Qi, J. Jia, Y. Yang, W. Duan, W.-X. Li, X. Bao, S. B. Zhang and Q.-K. Xue, *Proc. Natl. Acad. Sci. USA*, 2007, **104**, 920–9208.
9 X. Liu, C.-Z. Wang, M. Hupalo, H.-Q. Lin, K.-M. Ho and M. C. Tringides, *Phys. Rev. B*, 2014, **89**, 041401(R).
10 Y. Qi, X. Ma, P. Jiang, S. H. Ji, Y. S. Fu, J. F. Jia, Q. K. Xue and S. B. Zhang, *Appl. Phys. Lett.*, 2007, **90**, 013109.
11 J. Kim, S. Qin, W. Yao, Q. Niu, M. Y. Chou and C. K. Shih, *Proc. Natl. Acad. Sci. USA*, 2010, **107**, 12761–12765.
12 R. Otero, A. L. Váquez de Parga and R. Miranda, *Phys. Rev. B*, 2002, **66**, 115401.
13 M. Becker and R. Berndt, *Appl. Phys. Lett.*, 2010, **96**, 033112.
14 M. Hupalo, S. Kremmer, V. Yeh, L. Berbil-Bautista, E. Abram and M. C. Tringides, *Surf. Sci.*, 2001, **493**, 526–538.

15 C. M. Wei and M. Y. Chou, *Phys. Rev. B*, 2002, **66**, 233408.

16 L. Wang, X. C. Ma, P. Jiang, Y. S. Fu, S. H. Ji, J. F. Jia and Q. K. Xue, *J. Phys.: Cond. Matter*, 2007, **19**, 306002.

17 P. S. Kirchmann, M. Wolf, J. H. Dil, K. Horn and U. Bovensiepen, *Phys. Rev. B*, 2007, **76**, 075406.

18 Y.-F. Zhang, J.-F. Jia, T.-Z. Han, Z. Tang, Q.-T. Shen, Y. Guo, Z. Q. Qiu and Q.-K. Xue, *Phys. Rev. Lett.*, 2005, **95**, 096802.

19 P. Czoschke, H. Hong, L. Basile and T. C. Chiang, *Phys. Rev.Lett.*, 2004, **93**, 036103.

20 T. Miller, M. Y. Chou and T.-C. Chiang, *Phys. Rev. Lett.*, 2009, **102**, 236803.

21 M. Nonnnmacher, M. P. Oboyle and H. K. Wickramasinghe, *Appl. Phys. Lett.*, 1991, **58**, 2921–2923.

22 S. A. Burke, J. M. LeDue, Y. Miyahara, J. M. Topple, S. Fostner and P. Grütter, *Nanotechnology*, 2009, **20**, 264012.

23 J. L. Neff, P. Milde, C. Pérez León, M. D. Kundrat, C. R. Jacob, L. Eng and R. Hoffmann-Vogel, *ACS Nano*, 2014, **8**, 3294–3301.

24 C. Pérez León, H. Drees, S. M. Wippermann, M. Marz and R. Hoffmann-Vogel, *J. Phys. Chem. Lett.*, 2016, **7**, 426–430.

25 R. Feng, E. H. Conrad, M. C. Tringides, C. Kim and P. F. Miceli, *Appl. Phys. Lett.*, 2004, **85**, 3866–3868.

26 C. Barth and C. R. Henry, *Phys. Rev. Lett.*, 2007, **98**, 136804.

27 H.-Q. Mao, N. Li, X. Chen and Q.-K. Xue, *Chin. Phys. Lett.*, 2012, **29**, 066802.

28 S. Sadewasser and M. C. Lux-Steiner, *Phys. Rev. Lett.*, 2003, **91**, 266101.

29 I. B. Altfder, K. A. Matveev and D. M. Chen, *Phys. Rev. Lett.*, 1997, **78**, 2815–2818.

30 Y. Jia, B. Wu, C. Li, T. L. Einstein, H. H. Weitering and Z. Zhang, *Phys. Rev. Lett.*, 2010, **105**, 066101.

31 J. Kim, C. Zhang, J. Kim, H. Gao, M.-Y. Chou and C.-K. Shih, *Phys. Rev. B*, 2013, **87**, 245432.

32 Y. Jia, B. Wu, H. H. Weitering and Z. Zhang, *Phys. Rev. B*, 2006, **74**, 035433.

33 S. Pan, Q. Liu, F. Ming, K. Wang and X. Xiao, *J. Phys.: Cond. Matter*, 2011, **23**, 485001.

34 M. Liu, Y. Han, L. Tang, J.-F. Jia, Q.-K. Xue and F. Liu, *Phys. Rev. B*, 2012, **86**, 125427.

# Near-Equilibrium Measurement of Quantum Size Effects Using Kelvin Probe Force Microscopy

Thomas Späth,[†,‡] Matthias Popp,[†] Carmen Pérez León,*,[†] Michael Marz,[†] and Regina Hoffmann-Vogel[†,¶]

*Physikalisches Institut, Karlsruhe Institute of Technology (KIT), Wolfgang-Gaede-Str. 1, 76128 Karlsruhe, Germany*

E-mail: carmen.perez.leon@kit.edu

## Contents:

I- Simple model for calculating the Fermi energy

II- Atomically resolved SFM image of a Pb island. Figure S1

III- Statistical averaging methods. Figures S2, S3, S4

IV- Extra data at large island heights. Figure S5

V- Fitting parameters. Table S1

*To whom correspondence should be addressed
[†]Physikalisches Institut, Karlsruhe Institute of Technology (KIT), Wolfgang-Gaede-Str. 1, 76128 Karlsruhe, Germany
[‡]Current address: Department of Materials Science, Darmstadt University of Technology, Jovanka-Bontschits-Str. 2, 64287 Darmstadt, Germany
[¶]Institut für Angewandte Physik, Karlsruhe Institute of Technology (KIT), Wolfgang-Gaede-Str. 1, 76128 Karlsruhe, Germany

# I- Simple model for calculating the Fermi energy

In this section, we will calculate the Fermi energy for thin films. The electrons in the Pb islands can move almost freely in the lateral direction in the island. In the direction perpendicular to the surface, the electrons are confined in the potential well formed between the vacuum-metal interface on one side and the band gap of the Si surface (the substrate) at the other side resulting in the quantization of the energy levels[1]. Thus, the number of states in $x$- and $y$-direction is quasi-continuous whereas in the $z$-direction the wavevector component $k_z$ is quantized. Consequently, the Fermi sphere of allowed states is reduced to a discrete number of Fermi discs with a constant $k_z$ value.

We will use the solution of the Schrödinger's equation for a potential well with a thickness $a$ and a lateral extension of $L$, where $L >> a$. The wavefunction is separable into three parts for the three spatial directions:

$$\Psi(x, y, z) = \Psi_1(x)\Psi_2(y)\Psi_3(z), \tag{S1}$$

The solution in $x$- and $y$-direction are plane waves, whereas in $z$-direction, this is given by the solution of the one-dimensional Schrödinger's equation for the potential barrier of the well.

The number of available electronic states is determined by the area of the Fermi discs (Figure 1a). The state with $k_z = 0$, i.e. $n_z = 0$, is a non-physical solution of the equation. The state with $k_z$ is equal to the state $-k_z$ and is therefore not counted, only $n_z > 0$ values are valid. For $n_z = 1$, the area is given by:

$$A_1 = \pi(k_F^2 - k_{z,1}^2) \tag{S2}$$

and the number of states by:

$$N_1 = \left(\frac{L}{2\pi}\right)^2 \cdot \pi(k_F^2 - k_{z,1}^2). \tag{S3}$$

For an arbitrary $\ell$, being $\ell < n_z$ and $2n_z$ the number of available Fermi discs, the number of available states is then given by:

$$N_{n_z} = \left(\frac{L}{2\pi}\right)^2 \cdot \pi \sum_{\ell=1}^{n_z} (k_F^2 - k_{z,\ell}^2). \tag{S4}$$

Not to forget that the total number of states is twice as large due to the two possible spin orientations, thus

$$N = 2 \cdot N_{n_z}. \tag{S5}$$

On the other hand, the number of electrons in the island is $N = \rho V = \rho L^2 a$, $\rho$ being the electron density. Resolving for the Fermi wavevector $k_F$, we obtain:

$$k_F = \sqrt{\frac{2\pi \rho a + \sum_{\ell=1}^{n_z} k_{z,\ell}^2}{n_z}}, \tag{S6}$$

and for the Fermi energy:

$$E_F = \frac{\hbar^2}{2m} \frac{2\pi \rho a + \sum_{\ell=1}^{n_z} k_{z,\ell}^2}{n_z}. \tag{S7}$$

Both Equations S6 and S7 describe a set of curves for varying $n_z$. The Fermi energy of the island is thus piecewise defined by the state that has the lowest energy for a given island thickness.

## I.1 Infinite potential well:

Using the boundary conditions of an infinite potential well, the wavefunctions must be zero at the borders of the well:

$$k_z \cdot a = n_z \pi \quad \rightarrow \quad a = n_z \cdot \frac{1}{2} \lambda_z, \tag{S8}$$

where $n_z$ is an integer. Consequently, the island's thickness $a$ must be a multiple of the half of the wavelength $\lambda_z$. In the $k_z$-direction, the distance between states, $\pi/a$, decreases by increasing the thickness of the island $a$.

Using Equation S8, $k_z = n_z \pi/a$, with a distance between states of $\pi/a$, Equation S4 can be simplified to

$$N_{n_z} = \left(\frac{L}{2\pi}\right)^2 \cdot \pi \sum_{\ell=1}^{n_z}(k_F^2 - k_{z,\ell}^2) = \left(\frac{L}{2\pi}\right)^2 \cdot \left(\pi n_z k_F^2 - \pi \left(\frac{\pi}{a}\right)^2 \sum_{\ell=1}^{n_z} \ell^2\right). \tag{S9}$$

The sum is equal to

$$\sum_{\ell=1}^{n_z} \ell^2 = \frac{1}{6}n_z(n_z+1)(2n_z+1), \tag{S10}$$

such that

$$k_F = \sqrt{\frac{2\pi \rho a}{n_z} + \frac{\pi^2}{6}\frac{(n_z+1)(2n_z+1)}{a^2}}. \tag{S11}$$

The value of $k_F$ decreases for small $a$ proportional to $1/a$, has a minimum and increases again for large $a$ proportional to $\sqrt{a}$.

We now write the sample thickness $a$ as a function of the bulk Fermi wavelength $\lambda_{F,bulk}$ as $a = r\lambda_{F,bulk}$, where $r = n_z/2$. In addition, we divide $k_F$ by its bulk value and get:

$$\frac{k_F}{k_{F,bulk}} = \sqrt{\frac{4}{3}\frac{r}{n_z} + \frac{1}{24}\frac{(n_z+1)(2n_z+1)}{r^2}}, \tag{S12}$$

that is $\rho$ independent. Using that $E_F = (\hbar k_F)^2/(2m)$, we obtain for the Fermi energy:

$$\frac{E_F}{E_{F,bulk}} = \frac{k_F^2}{k_{F,bulk}^2} = \frac{4}{3}\frac{r}{n_z} + \frac{1}{24}\frac{(n_z+1)(2n_z+1)}{r^2}. \tag{S13}$$

These equations, as in the case of $k_F$ and $E_F$ in Equations S6 and S7, describe a set of curves for varying $n_z$. The Fermi energy of the island is piecewise defined by the state that has the lowest energy for a given island thickness. The points where the curves intersect occur roughly at the integer multiples of $\lambda_{F,bulk}/2$. For large $n_z$, the minima of the curves converges versus the bulk value $E_{F,bulk}$. However, the ratio $E_F/E_{F,bulk}$ decays rather than oscillates and does not reproduce our experimental results.

## I.2 Finite potential well:

Using instead a finite potential well, the wavefunction can leak out of the potential well, and $k_z$ can be arbitrarily small as long as the potential barrier of the well also remain small. Inside the well, the wavefunction is described by a sinusoidal function, whereas outside the well it is described by an exponential decaying wavefunction, with the decay exponent $\kappa$. At the borders of the well, the two functions must be differentiable. This is achieved by

$$\kappa = k_{z,n_z} \tan\left(\frac{k_{z,n_z} L}{2}\right) \tag{S14}$$

for symmetric solutions, where $y$ is similar to a cosine and by

$$\kappa = -k_{z,n_z} \cot\left(\frac{k_{z,n_z} L}{2}\right) \tag{S15}$$

for asymmetric solutions, where $y$ is similar to a sine. In addition we have

$$\kappa^2 + k_{z,n_z}^2 = \frac{2m V_0}{\hbar^2}, \tag{S16}$$

where $V_0$ is the potential barrier of the well. $k_{z,n_z}$ is calculated numerically with these equations. The result for $E_F/E_{F,bulk}$ is the piecewise defined curve shown in Figure 1b of the main manuscript. This graph shows the resulting Fermi energy for the different states defined by different $n_z$. The main features such as a dampen oscillatory behavior and an oscillation wavelength of $\lambda_{F,bulk}/2$ are well described by this model. The decay, however, is given by $1/r^2$, see Equation S13 instead of $1/r$ which is closer to our experimental observations. Although the basic physics can be understood from the considerations made here, the result does not describe the experimental data in detail.

## II - Atomically resolved SFM image on a Pb island

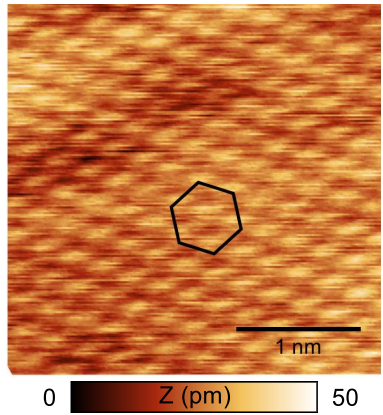

Figure S1: Atomically resolved SFM topographic image obtained in the dynamic mode with FM detection. Flat-top mesa shaped (111)-oriented Pb islands grow on Si(111) surrounded by a wetting layer. The black hexagon highlights the atomic arrangement of the Pb(111) surface. The image has been drift-corrected.

**Imaging parameters:** $(3 \times 3)$ nm$^2$; $\Delta f = -57$ Hz, $A = 8$ nm, $k = 50$ N/m, $f_0 = 280059$ Hz, $\gamma = 7.3$ fN$\sqrt{\text{m}}$. Pt-Ir coated tip.

## III - Statistical averaging methods

Our SFM images in Figure 2 of the main manuscript show small Pb islands, several of them partly extending over more than one Si(111) terrace or having internal height differences. We analyzed the islands using a statistical averaging method that we introduce in the following for a Pb island having one particular height, and, accordingly, one defined local work function (LWF). This method has been proven to be suitable also for islands spreading over several Si terraces. First, the image of a single Pb island was cut from an overview image (as shown in Figure S2a). A histogram of the measured heights was generated for the cut image (see

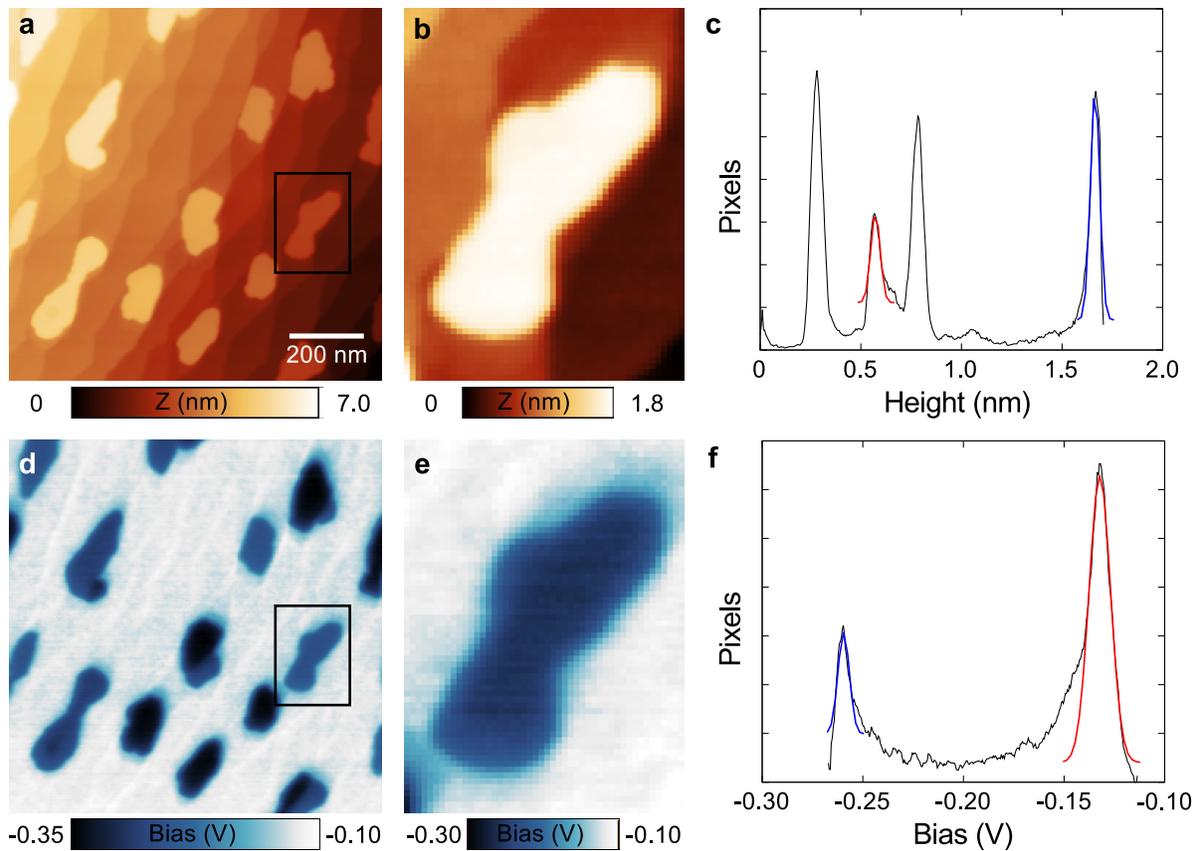

Figure S2: a) Overview SFM image of small Pb islands. b) First, the image of a single island was cut. c) Then a histogram of the measured heights of the cut image was generated. The peaks corresponding to the island surface and to the terrace where the island sits were identified and fitted by Gaussian functions (in blue and red respectively). The terraces are covered by the wetting layer. A similar procedure was applied to the Kelvin image: d) Overview Kelvin image of Pb islands, e) cut image of an island, and f) corresponding histogram generated from e.

Figure S2b,c). Figure S2a is the same image as Figure 2a, but leveled such that the terraces and the island's surfaces are horizontally flat. This leveling is useful for the analysis with histograms. The peak in the histogram corresponding to the height of the island and the peak corresponding to the height of the terrace where the island sits were identified and fitted by a Gaussian function, blue and red fits respectively. This height corresponds to the height of the terrace covered by the wetting layer (WL). Since, as we explained in the main manuscript our strategy is to analyze the data using the WL as reference, the difference between the two centers of the Gaussian functions was taken to be the island height. The

S7

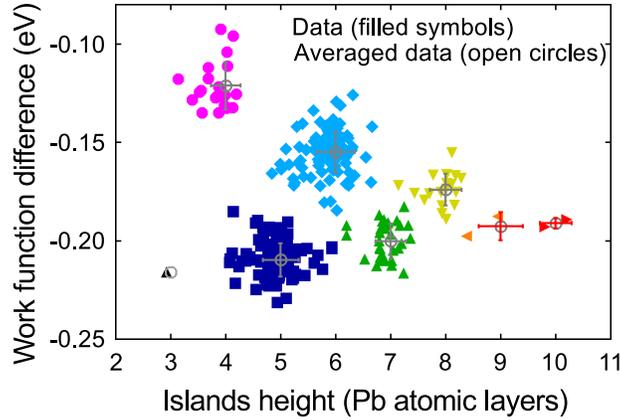

Figure S3: Data results after the statistical analysis of many experiments performed with one single Pt/Ir-coated tip. They correspond to the statistical spread of the data in Figure 4 of the main manuscript (here shown in grey). Different island heights and distinct local work function differences are clearly identified. The local height of the islands is given in atomic layers measured from the WL.

same procedure was performed for the Kelvin images (Figure S2d–f). In this way the LWF values are referred to the WL, and we obtain tip-independent results.

Figure S3 shows the data results after the statistical analysis of many experiments performed with one single Pt/Ir-coated tip. The height of the islands with respect to the WL is given in Pb atomic layers. In the graph, the data resulting from a particular island thickness can be easily identified. For every height, the data were averaged and plotted with error bars (in grey in Figure S3). These averaged results correspond to the data in Figure 4 of the main manuscript.

The results obtained from different several measurement series using different tips are shown in Figure S4. It is noticeable that data obtained with different tips align well with each other, confirming that we indeed obtain tip-independent results.

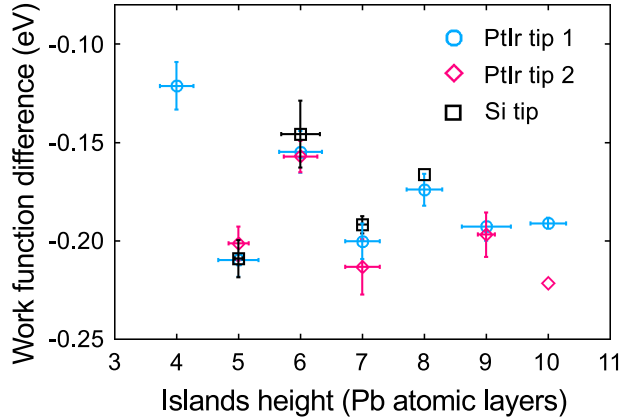

Figure S4: Quantum oscillations in the local work function of ultra-thin Pb islands on Si(111) as a function of the island height measured from the WL obtained with diverse tips. The averaged data with error bars are plotted. Data obtained with different tips align well with each other, proving that we indeed obtain tip-independent results.

# IV - Extra data at large island heights

In addition to the data shown in the main paper, we have studied thicker islands. As was discussed in the main text, the amplitude of the bilayer oscillation of the LWF with respect to the film thickness decays with for increasing island height. The oscillation is additionally modulated by a quantum beating pattern that causes a periodical phase slip every nine to ten atomic layers. This implies an increase of the oscillation amplitude for islands higher than the investigated range in the main manuscript (10 atomic layers). The purpose of this study is to confirm the existence of a beating pattern and to precisely locate the island height for which the knot of the quantum beating pattern is observed.

Large island heights were obtained by annealing the samples at room temperature after lead evaporation for a defined time (between 15 and 90 min). Subsequently, the sample was cooled down with liquid nitrogen for SFM measurements. Afterwards, the samples were repeatedly warmed up and cooled down again for further measurements. The results of this new set of experiments with combined SFM and Kelvin probe force microscopy (KPFM) were analyzed using different methods than for the case of lower island heights. In this case, line profiles over the images were used to determine the island heights and their Kelvin voltage. The values obtained for the thinner islands or part of the islands below 10 atomic layers in height measured from the WL mostly agree with the data shown in Figure 4 of the main manuscript. Figure S5 presents the additional data for islands up to 14 atomic layers measured from the WL combined with the results of Figure 4. An increase of the quantum oscillation amplitude and a reversal of maxima and minima beyond 10 atomic layers is clearly observed, the knot being placed between $9 - 10$ atomic layers from the WL.

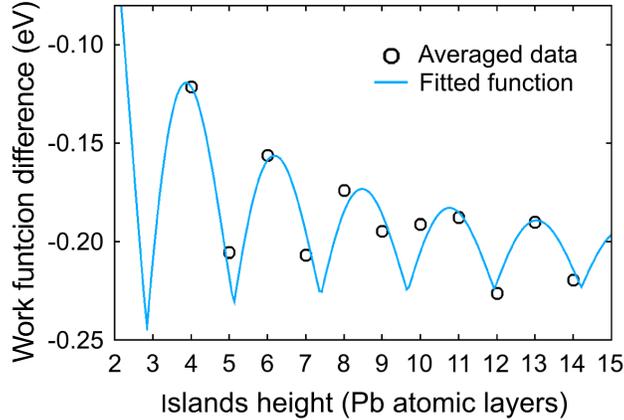

Figure S5: Quantum oscillations in the local work function of Pb islands on Si(111) with island heights up to 14 atomic layers as a function of the island height measured from the WL obtained with SFM/KPFM. Results obtained from an additional set of experiments including data at larger island heights combined with the data from Figure 4 of the main manuscript. In addition to the damped even-odd oscillating with increasing island height, these data show the quantum beating pattern: At nine atomic layers the amplitude of the oscillation vanishes and increases again for larger film thicknesses. Data fitted with function of Equation 6 of the main manuscript and Equation S17 of Supporting Information ($\alpha = 1.02$).

We observe that the value of the LWF difference obtained on some samples of this additional set of experiments for obtaining higher islands is shifted by few mV towards more negative values with respect to the other samples and to results in Figure 4 of the main manuscript. These shifts appear for samples that have been exposed to long annealing times at room temperatures (at least 90 min). We also observed that if the samples were kept long enough at low temperatures (several days), the LWF shifted back to the original values. It is important to state that these shifts affect a particular measurement series as a whole, and not only data for particular islands' heights. Here, we are interested in the decay of the oscillation amplitude as a function of the film thickness, and in the position of the knot of the beating pattern, i.e. the phase of the beating pattern. Thus, we have chosen only similar data for averaging, ensuring that the amplitude of the oscillation is not markedly influenced by the shifts.

We tentatively explain the shifts of the LWF differences to be due to variations in the LWF of our reference, the wetting layer. The WL is a dynamic system that transfers material collectively to islands even at low temperatures.[2] This material comes from unstable islands and the WL itself.[2–4] The thickness of the WL was reported to be between $1-3$ atomic layers depending on the growth temperature,[3,5] but it is now generally agreed that it is around one atomic layer.[6,7] The density of the WL is 22% larger than the metallic Pb(111) density.[4] This is possible because the Si substrate facilitates the Pb atoms to move closer than their average distance in bulk Pb.[4] It has been pointed out that fluctuations exist within the WL, suggesting variations in the WL density.[8] We presume that the long annealing at room temperature for obtaining large islands affects the structure of the WL. Due to coarsening, the WL rearranges or looses atoms that are transferred to larger islands. This possibly has as consequence a decrease in density of the WL. This less dense WL would have a higher LWF than the usual dense one, and the LWF difference between the islands and the WL would be larger, i.e. more negative, like the shift that we observe in our samples annealed for long times.

## V - Fiting parameters

The phenomenological damped sinusoidal function reported in the main manuscript as Equation 6:

$$f(N) = \frac{A|\sin Nk_F d_0 + \phi| + B}{N^\alpha} + C, \tag{S17}$$

has been used to describe the periodicity and decay of the oscillation of the LWF as a function of the number of atomic layers measured from the WL ($N$), as well as the beating pattern. In Equation S17, $A$ (amplitude parameter), $B$ (amplitude offset), $C$ (constant offset), $\phi$ (phase shift factor that will be dependent on the interface properties of the film), and $\alpha$ (the decay exponent) are $N$-independent constants,[9,10] and $k_F$ is the Fermi wave vector measured from the zone boundary (see explanation below).[11]

Table S1: Free parameters obtained after fitting the experimental data with the damped sinusoidal function given in Equation 6 of the main manuscript, here Equation S17.

|   | Fig. 4 | Fig. S5 |
| --- | --- | --- |
| A | $0.595 \pm 0.200$ eV | $0.473 \pm 0.235$ eV |
| B | $-0.06 \pm 0.08$ eV | $-0.07 \pm 0.09$ eV |
| C | $-0.210 \pm 0.010$ eV | $-0.218 \pm 0.010$ eV |
| $\alpha$ | $1.26 \pm 0.20$ | $1.02 \pm 0.30$ |
| $\varphi$ | $2.62 \pm 0.03$ | $2.35 \pm 0.06$ |

To evaluate the decay exponent of the oscillations, we have fitted this equation to our data of Figure 4 and Figure S5. The values obtained for the free parameters are presented in Table S1. The value of $\alpha$ is larger than 1 for the fit of the data in Figure 4 due to the fitting only up to the first beating pattern. If we use data up to 14 atomic layers, the value gets closer to the expected $\alpha = 1$.

Using $\lambda_F$ and $d_0$ given in the main text for free electrons, we obtain $k_F d_0 = 1.44\pi$. With these values, Equation S17 has a period of less than one atomic layer and cannot be observed experimentally due to the discrete nature of the atomic layer structure of the films. The number of oscillations of the fit function is better described by the Fermi wave vector measured from the zone boundary, here $k_F d_0 = 0.44\pi$ and the oscillation period is around 2.

# References

(1) Milun, M.; Pervan, P.; Woodruff, D. P. *Rep. Prog. Phys.* **2002**, *65*, 99–141.

(2) Man, K. L.; Tringides, M. C.; Loy, M. M. T.; Altman, M. S. *Phys. Rev. Lett.* **2008**, *101*, 226102.

(3) Hupalo, M.; Kremmer, S.; Yeh, V.; Berbil-Bautista, L.; Abram, E.; Tringides, M. C. *Surf. Sci.* **2001**, *493*, 526–538.

(4) Hershberger, M. T.; Hupalo, M.; Thiel, P. A.; Wang, C. Z.; Ho, K. M.; Tringides, M. C. *Phys. Rev. Lett.* **2014**, *113*, 236101.

(5) Qi, Y.; Ma, X.; Jiang, P.; Ji, S. H.; Fu, Y. S.; Jia, J. F.; Xue, Q. K.; Zhang, S. B. *Appl. Phys. Lett.* **2007**, *90*, 013109.

(6) Wang, L.; Ma, X. C.; Jiang, P.; Fu, Y. S.; Ji, S. H.; Jia, J. F.; Xue, Q. K. *J. Phys.: Cond. Matter* **2007**, *19*, 306002.

(7) Kim, J.; Zhang, C.; Kim, J.; Gao, H.; Chou, M.-Y.; Shih, C.-K. *Phys. Rev. B* **2013**, *87*, 245432.

(8) Hattab, H.; Hupalo, M.; Hershberger, M. T.; von Hoegen, M. H.; Tringides, M. C. *Surf. Sci.* **2016**, *646*, 50–55.

(9) Wei, C. M.; Chou, M. Y. *Phys. Rev. B* **2002**, *66*, 233408.

(10) Czoschke, P.; Hong, H.; Basile, L.; Chiang, T.-C. *Phys. Rev. B* **2005**, *72*, 075402.

(11) Miller, T.; Chou, M. Y.; Chiang, T.-C. *Phys. Rev. Lett.* **2009**, *102*, 236803.